\documentclass[12pt,preprint]{aastex}

\def\ltsim{\raisebox{-.5ex}{$\;\stackrel{<}{\sim}\;$}}
\def\gtsim{\raisebox{-.5ex}{$\;\stackrel{>}{\sim}\;$}}

\shortauthors{Crenshaw \& Kraemer}
\shorttitle{Feedback in AGN}

\begin{document}

\title{Feedback from Mass Outflows in Nearby Active Galactic Nuclei I. UV and
X-ray Absorbers}

\author{D.M. Crenshaw\altaffilmark{1}
and S.B. Kraemer\altaffilmark{2}}

\altaffiltext{1}{Department of Physics and Astronomy, Georgia State 
University, Astronomy Offices, One Park Place South SE, Suite 700,
Atlanta, GA 30303; crenshaw@chara.gsu.edu}

\altaffiltext{2}{Institute for Astrophysics and Computational Sciences,
Department of Physics, The Catholic University of America, Washington, DC
20064}

\begin{abstract}

We present an investigation into the impact of feedback from outflowing UV and
X-ray absorbers in nearby ($z < 0.04$) AGN. From studies of the kinematics,
physical conditions, and variability of the absorbers in the literature, we
calculate the possible ranges in total mass outflow rate ($\dot{M}_{out}$) and
kinetic luminosity ($L_{KE}$) for each AGN, summed over all of its absorbers.
These calculations make use of values (or limits) for the radial locations of
the absorbers determined from variability, excited-state absorption, and other
considerations. From a sample of 10 Seyfert 1 galaxies with detailed
photoionization models for their absorbers, we find that 7 have sufficient
constraints on the absorber locations to determine $\dot{M}_{out}$ and $L_{KE}$.
For the low-luminosity AGN NGC~4395, these values are low, although we do not
have sufficient constraints on the X-ray absorbers to make definitive
conclusions. At least 5 of the 6 Seyfert 1s with moderate bolometric
luminosities ($L_{bol} = 10^{43} - 10^{45}$ ergs s$^{-1}$) have mass outflow
rates that are 10 -- 1000 times the mass accretion rates needed to generate
their observed luminosities, indicating that most of the mass outflow originates
from outside the inner accretion disk. Three of these (NGC~4051, NGC~3516, and
NGC~3783) have $L_{KE}$ in the range 0.5 -- 5\% $L_{bol}$, which is the range
typically required by feedback models for efficient self-regulation of
black-hole and galactic bulge growth. At least 2 of the other 3 (NGC~5548,
NGC~4151, and NGC~7469) have $L_{KE} \gtsim 0.1\% L_{bol}$, although these
values may increase if radial locations can be determined for more of the
absorbers. We conclude that the outflowing UV and X-ray absorbers in
moderate-luminosity AGN have the potential to deliver significant feedback to
their environments.

\end{abstract}

~~~~~

\section{Introduction}

Active galactic nuclei (AGN) are fed by accretion of matter onto supermassive
black holes (SMBHs), generating huge amounts of radiation from very small
volumes. In addition to radiative feedback (e.g., Ciotti, Ostriker, \& Proga
2010), AGN provide feedback via mass outflows of ionized gas into their
environments, which are thought to play a critical role in the formation of
large-scale structure in the early Universe (e.g., Scannapieco \& Oh 2004; Di
Matteo et al. 2005), chemical enrichment of the intergalactic medium (e.g.,
Khalatyan et al. 2008), and self-regulation of SMBH and galactic bulge growth
(e.g., Hopkins et al. 2005). For example, the currently popular explanation for
the relation between the SMBH mass and the stellar velocity dispersion in the
bulge, the $M_{BH} - \sigma_{*}$ relation (Gebhardt et al. 2000; Ferrarese \&
Merritt 2000), is that AGN feedback results in evacuation of gas from the bulge,
quenching of star formation, and a halt to the growth of the SMBH and bulge.
However, we have very little information on the validity of this explanation,
the frequency and magnitude of mass outflows from AGN, or the detailed physical
mechanisms of feedback.

Mass outflows from AGN arise in two principal sources: radio jets and ``AGN
winds''. Most feedback models have concentrated on the former, because jets are
very powerful and are clearly seen to impact their host galaxies and
extragalactic environments. However, jets are narrowly focused, and radio-loud
AGN with strong jets occur in only 5 -- 10\% of the AGN population (Rafter et
al. 2009, and references therein). Thus, it is important to consider the impact
that AGN winds have on their environments. These winds are often revealed
through UV and X-ray absorption lines that are blueshifted with respect to their
host galaxies, with typical outflow velocities up to 2000 km s$^{-1}$ in Seyfert
1 galaxies (Crenshaw, Kraemer, \& George 2003a) and potentially much higher
velocities in quasars (Ganguly \& Brotherton 2008), especially in broad
absorption-line (BAL) quasars (with maximum outflow velocities between 3000 and
25,000 km s$^{-1}$, Gibson et al. 1999). Winds in another form have also been
detected as outflows of emission-line gas in the narrow (emission) line regions
(NLRs) of nearby AGN, with outflow velocities up to $\sim$1500 km s$^{-1}$ on
scales of hundreds of parsecs (Crenshaw et al. 2010; Fischer et al. 2011).

Although AGN feedback is usually discussed in terms of high-luminosity quasars
interacting with their environments at high redshifts, it is useful to explore
the impact of winds from nearby AGN at moderate luminosities. In particular,
Seyfert 1 galaxies, with bolometric luminosities L$_{bol} $\ltsim$ 10^{45}$ erg
s$^{-1}$ cm$^{-2}$, are bright enough in apparent magnitude for high-resolution
spectroscopy to study the detailed physics of their winds. This pursuit is
important for gauging the importance of winds from moderate-luminosity AGN in
general, and for understanding the mechanisms of feedback in more luminous
AGN at higher redshifts.

In this paper, we concentrate on outflowing UV and X-ray absorbers, seen as
kinematic components of blueshifted absorption lines in the spectra of Seyfert 1
galaxies. A large number of high-resolution UV and X-ray spectra suitable for
this purpose have been obtained over the past couple of decades with the {\it
Hubble Space Telescope} ({\it HST}), {\it Far Ultraviolet Spectroscopic
Explorer} ({\it FUSE}), {\it Chandra X-ray Observatory} ({\it CXO}), and {\it
X-ray Multi-Mirror Mission} ({\it XMM-Newton}). The absorbers have been
characterized in great detail by ourselves and others with the use of
multi-epoch observations and detailed photoionization models. In this paper, we
use the published results to estimate the magnitude of feedback from UV and
X-ray absorbers, by determining (or placing limits on) the total mass outflow
rate ($\dot{M}_{out}$) and kinetic luminosity ($L_{KE}$ ) in individual AGN. In
a
subsequent paper, we will examine the NLR outflows in Seyfert galaxies and their
importance for AGN feedback.

\section{Sample}

In order to quantify the feedback from outflowing absorbers in nearby AGN, we
need UV and X-ray spectra at high spectral resolutions to isolate and measure
the kinematic components of absorption. We apply this criterion by selecting
studies with {\it HST} UV observations over the 1150 -- 3200 \AA\ range at
velocity resolutions of 7 -- 30 km s$^{-1}$ (FWHM) with the Goddard
High-Resolution Spectrograph (GHRS), Space Telescope Imaging Spectrograph
(STIS), and Cosmic Origins Spectrograph (COS). We can also use spectra from {\it
FUSE} in the range 900 - 1200 \AA\ at a resolution of $\sim$15 km s$^{-1}$
(FWHM). To obtain the highest possible spectral resolutions in X-rays, we make
use of studies with observations primarily from the grating spectrometers on
{\it CXO} in conjunction with the HETG (0.4 -- 10 keV coverage, $\geq$300 km
s$^{-1}$ resolution in the regions of interest) or LETG (0.1 -- 3 keV coverage,
$\geq$150 km s$^{-1}$
resolution) gratings, and from the {\it XMM-Newton} Reflection Grating
Spectrometer (RGS) (0.25 -- 2.5 keV coverage, $\geq$400 km s$^{-1}$ resolution).

Our second criterion is that there must be detailed photoionization models for
both UV and X-ray absorbers in each AGN in the literature, so that we have the
measurements needed for feedback determinations. The models are based on
measured ionic column densities and provide the ionization parameters and
hydrogen column densities that characterize the physical conditions in the gas.
In the UV, one must correct for partial covering of the background emission
(continuum, broad line, etc.) by the absorbers to avoid underestimating the
column densities (Arav et al. 2002, 2003; Crenshaw et al. 2003a).

We use the dimensionless ionization parameter $U$, which is the density of
photons with energies $\geq$ 13.6 eV divided by the number density of hydrogen
atoms at the illuminated face of the slab:
\begin{equation}
{\rm U} = \int_{\nu_{0}}^{\infty}{{\rm L_{\nu}} h\nu\over{4\pi r^{2} n_H c}} 
d\nu.
\end{equation}
In studies where the ionization parameter $\xi$ ($=L_{ion}/n_Hr^2$) is given, we
use the conversion log$(U) = $ log($\xi) - 1.5$ based on a typical Seyfert 1
spectral energy distribution (SED), which we take to be several joined power
laws of the form $F_{\nu} \propto \nu ^{-\alpha}$, with $\alpha = 1$ below 13.6
eV, $\alpha = 1.4$ over the range 13.6 eV $<$ h$\nu$ $<$ 1000 eV, and $\alpha =
0.7$ above 1000 eV (Kraemer et al. 2001).

The other model parameter that we use is the total hydrogen density $N_H$ $=
N_{HI} + N_{HII}$ in units of cm$^{-2}$. $N_H$ is sensitive to the
elemental abundances used in the photoionization models, which are
specified by the studies that present these models. In most cases, solar
abundances (Asplund et al. 2009) are used, but there are a few
notable exceptions of higher metallicity outflows (see Arav et al. 2007; Fields
et al. 2007).

Our third criterion is that we restrict our sample to apparently bright AGN with
broad emission lines at redshifts $z < 0.04$ that are suitable for high
resolution spectroscopy. We do not include ultrafast outflows (with
outflow velocities $>$ 10,000 km s$^{-1}$, Tombesi et al. 2010), as the nature
and global covering factor of these absorber outflows are not well understood.
We note that neither BALs (e.g., Gibson et al. 2009) nor very high velocity UV
absorbers (e.g., Hamann et al. 2011) have been detected in AGN at these low
redshifts.

The above criteria result in a sample of 10 nearby Seyfert 1 galaxies
listed in Table 1. We note that NGC~4051 and Akn~564 are often classified as
narrow-line Seyfert 1 galaxies, with FWHM(broad H$\beta$) $<$ 2000 km s$^{-1}$,
and NGC 4395 is often called a ``dwarf Seyfert 1'' due to its low luminosity.

In Table 1, we give some of the fundamental parameters for each AGN in our
sample. From the literaure, we give the estimated radii of the broad line
regions (BLRs) in light days from reverberation mapping of the broad C~IV and
H$\beta$ emissions, $r_{CIV}$ and $r_{H\beta}$ respectively (Peterson et al.
2004).
For each AGN, we also list the derived mass of the SMBH, $M_{BH}$, and the
monochromatic luminosity at 5100 \AA, $\lambda$L$_{\lambda}(5100)$, both
obtained from the listed reference. In addition to $M_{BH}$, we give the
bolometric luminosity L$_{bol}$ = 9.8 $\lambda$L$_{\lambda}(5100)$ (McLure \&
Dunlop 2004), Eddington ratio $L_{bol}/L_{Edd}$, and mass accretion rate needed
to generate the observed luminosity $\dot{M}_{acc} = L_{bol}/\eta c^ 2$, where
we assume $\eta = 0.1$ (Peterson et al. 1997).

\section{Analysis}
\subsection{Feedback Calculations}

To determine the mass outflow rate for each absorber, we use the equation
\begin{equation}
\dot{M}_{out} = 4 \pi r N_H \mu m_p C_g v_r
\end{equation}
(Crenshaw et al. 2003a), where $r$ is the absorber's radial location (i.e., its
distance from the central SMBH), $N_H$ is the hydrogen column density, $\mu$ is
the mean atomic mass per proton ($=$ 1.4 for solar abundances), $m_p$ is the
proton mass, $C_g$ is the global covering factor ($=$ 0.5), and $v_r$ is the
radial velocity centroid. The kinetic luminosity is then:
\begin{equation}
L_{K} = 1/2\dot{M}_{out}v_r^2 = 2 \pi r N_H \mu m_p C_g v_r^3
\end{equation}

The average global covering factor can be determined statistically from $C_g =
C_{los} f$, where $C_{los}$ is the average covering factor of the background
emission in the line of sight and $f$ is the fraction of AGN that show intrinsic
absorption (Crenshaw et al. 2003a). A number of studies have shown that $C_g
\approx 0.5$ for both UV (Crenshaw et al. 1999; Dunn et al. 2007) and X-ray
absorbers (Reynolds 1997; George et al. 1998; Winter 2010). Technically,
this factor could be lower by a factor of 2 to 3, because, according to unified
models, we see Seyfert 1s over a restricted range of viewing angles. On the
other hand, if it turns out that half of Seyfert 1s are completely covered and
the other half are not covered at all, we could increase $C_g$ by a factor of
$\sim$2 for the former. Thus, we keep $C_g = 0.5$, with the
understanding that this value could be off by a factor of $\sim$2 either way.

\subsection{Absorber Distances}

In order to determine $\dot{M}_{out}$ and $L_{KE}$ for each absorber,  we do
{\it not} assume continuous, radiatively driven outflows at constant velocity
(e.g., Blustin et al. 2005), as there is no evidence that this assumption is
valid. Thus, we must know the radial location $r$ of the absorber, which is the
most difficult parameter to determine in the above equations. Fortunately, we
can use absorption from excited states or absorption variability to determine
(or place limits on) $r$. As discussed in Crenshaw et al. (2003a), column
densities of excited levels populated by collisional excitation can be used to
determine the electron number density n$_e$ (and hence the hydrogen number
density n$_H$). If the absorption responds to a decrease in ionizing flux, one
can also determine n$_e$ and n$_H$ from the recombination time scale (Nicastro
et al. 1999). Photoionization models of the ionic column densities provide $U$,
and together with $n_H$ and the ionizing luminosity in photons, the radial
location $r$ via equation 1. If the absorption responds to an increase in
ionizing flux, $r$ can be determined directly from the ionization time scale
(Crenshaw et al. 2003a), which depends on the ionizing flux incident on the
cloud. In many cases, we only have upper limits on the time scale over which the
absorption lines vary, which yield lower limits on n$_H$ and upper limits on
$r$. On the other hand, if the absorbers do not respond to large changes in the
ionizing continuum, we can obtain upper limits to n$_H$ and lower limits to $r$.
The dominant uncertainites in these values come from uncertainties in the
photoionization model parameters; typical uncertainities are $\sim$0.3 in log
($r$) (Crenshaw et al. 2003, 2009; Kraemer et al. 2006).

We also use other clues to determine limits on $r$ for the absorbers. The size
of the BLR responsible for most of the broad C~IV emission, $r_{CIV}$, serves as
the absolute minimum for the radial location of the absorber, because in nearly
every case it has been shown that the depth of the C~IV absorption exceeds the
continuum flux, and thus this region must be at least partially covered
(Crenshaw et al. 2003a). As shown in Table 1, only 4 AGN in our sample have
direct determinations of $r_{CIV}$ via reverberation mapping (Peterson et al.
2004, 2005). They all have determinations of the size of the region responsible
for broad H$\beta$ emission, $r_{H\beta}$, which is known to be larger than
$r_{CIV}$ (Peterson et al. 2004, Vestergaard \& Peterson 2006). For the AGN in
Table 1, $r_{H\beta}$/$r_{CIV}$ $=$ 1.8 -- 2.7, whereas Netzer (2009)
suggests that this ratio in general is $\sim$3, based on scaling relations. To
be conservative, for the AGN without direct determinations of $r_{CIV}$ we use
$r_{CIV}$ $=$ $r_{H\beta}$/3.0 as the absolute minimum for the radial
location of the absorber. Uncertainites in the BLR sizes from reverberation
mapping are 0.1 to 0.2 in log ($r$) (Peterson et al. 2004). We note that the BLR
size does not provide a constraint on the X-ray absorbers, which lack
significant underlying broad-line emission.

Limits on $r$ can also be determined if there is evidence for or against the
absorber covering the NLR (Crenshaw et al. 2002, 2009), as long as there is an
estimate of the NLR size. In some cases, the relative locations of the absorbers
can be deduced from the photoionization models by finding, for example, that one
absorber cannot be shielded by another, and therefore must be inside of the
latter (Kraemer et al. 2002).

We can determine an absolute maximum for the radial location of an absorber,
based on the requirement that the thickness of the absorber cannot exceed its
distance from the SMBH: \begin{equation} r \leq \frac{L_{ion}}{N_H \xi}
\end{equation} (Blustin et al. 2005), where $L_{ion}$ is the ionizing luminosity
(27\% of L$_{bol}$ for our SED) and $\xi$ is determined from $U$ as previously
discussed. Uncertainties in this limit once again come from those in $U$ and
$N_H$, and are on the order of 0.3 in log($r$). This upper limit and the lower
limit from the BLR size provide extreme ranges to the absorber radial locations
that are only occasionally useful. Much tighter constraints come from absorption
from excited levels or variable absorption.

\section{Results}

We give a detailed account of the measurements from the literature that we
adopted for each AGN in the Appendix. We list these measurements and the derived
minimum and maximum $r$, $\dot{M}_{out}$, and $L_{KE}$ for each absorber in
Table 2, when these could be determined. Here we take a graphical look at some
of these values.

\subsection{Absorption Measurements}

In Figure 1, we show the full width at half-minimum (FWHM) of each absorption
component versus the radial-velocity centroid for the UV absorbers (as explained
in the Appendix, we do not have reliable FWHM for the X-ray absorbers). There is
no apparent correlation between the two parameters. The absorbers in these
Seyfert 1s span the velocity range $-$2000 to $+$ 200 km s$^{-1}$ with respect
to the systemic velocity of the host galaxy (with the exception of two X-ray
absorbers given in the Appendix) and the FWHM range is 20 to 500 km s$^{-1}$
(with one exception).

In Figure 2, we plot log ($U$) versus $v_r$ for both UV and X-ray absorbers. The
X-ray absorbers tend to have higher $U$, as expected. Again, there is no
correlation, despite a few claims in the literature of trends in individual AGN.
There is a huge range in ionization parameter: log($U$) $\approx$ $-$2 to 4.

In Figure 3, there appears to be a positive correlation between log ($U$) and
log ($N_H$). The lack of high-ionization columns at low column densities can be
explained by the sensitivity limits of current X-ray missions. It is not clear
why there are no large-column, low-ionization components detected in the UV --
this may represent a real physical constraint. There is a gap around log ($U$)
$=$ 0 in this plot, also seen in Figure 2. This gap may represent the manner
in which absorbers are typically identified, which is from the presence of C~IV
absorption in the UV, and O~IV or O~VIII absorption in the X-rays, and could
potentially be filled in with photoionization models of AGN observed by {\it
FUSE}, which provide access to O~VI at low redshift. To test this notion we ran
a photoionization model with log($U$) $=$ 0, log($N_H$) $=$ 21.0, and the
above SED, and found N(C~IV) $=$ 3.8 $\times$ 10$^{13}$ cm$^{-2}$, which can be
difficult to detect in {\it HST} spectra (Crenshaw et al. 1999), and N(O~VI) $=$
1.2 $\times$ 10$^{16}$ cm$^{-2}$, which would be strong in {\it FUSE} spectra
(Dunn et al. 2007). Note that the extreme point in Figure 3 at log($U$) $=$
-0.39, log($N_H$) $=$ 22.93 is the ``D+Ea'' component in NGC~4151 (see Table
3), which may result from a special line of sight near the edge of the NLR
bicone (Crenshaw \& Kraemer 2007).

We plot the values or limits for the radial location $r$ (in pc) along with
$v_r$ for each absorber in Figure 4. There is no apparent correlation between
the two. Compared to the UV absorbers, the X-ray absorbers tend to be
concentrated toward smaller $r$, but there is no preference in terms of $v_r$.
The main result from this figure is that the vast majority of UV and X-ray
absorbers lie between 0.01 and 100 pc from the central SMBH, outside of the BLR
and inside much of the classic NLR (i.e, in the inner NLR [Crenshaw \& Kraemer
2005] or the ``intermediate-line region'' [Crenshaw \& Kraemer 2007;
Crenshaw et al. 2009]). The issue of where the absorbers actually originate,
in contrast to where they are currently located, is
discussed in the next subsection and Section 5.

\subsection{Feedback Parameters}

In Table 2 we give the range in feedback parameters $\dot{M}_{out}$ and
$L_{KE}$ for each AGN from the minimum and maximum values summed over all
absorbers. We also give the ratio of outflow to accretion rate
$\dot{M}_{out}$/$\dot{M}_{acc}$ and the ratio of kinetic to bolometric
luminosity $L_{KE}/L_{bol}$ for each AGN. As discussed in the Appendix, we were
unable to obtain reliable limits for Mrk~279, Mrk~509, and Akn~564, due
primarily to the lack of constraints on radial locations for most of
their absorbers. Thus, we have feedback values for 7 of the 10 Seyfert 1
galaxies in our original sample. 

We plot the range in log($\dot{M}_{out}$/$\dot{M}_{acc}$) against $L_{bol}$ in
Figure 5. For 5 of the 7 Seyfert 1 galaxies in our sample, the mass outflow rate
exceeds the mass accretion rate by a factor of 10 to 1000 (NGC 7469 provides
only an upper limit and NGC 4395 is a low-luminosity Seyfert). Thus, the vast
majority of this type of outflow in moderate-luminosity AGN must originate
outside of the inner accretion disk, where most of the AGN's luminosity is
generated; otherwise, the inner accretion disk would likely quickly dissipate.
There may be a slight correlation of $\dot{M}_{out}$/$\dot{M}_{acc}$ with
$L_{bol}$, but better constraints and more data are needed, especially at
$L_{bol} = 10^{41} - 10^{43}$ ergs s$^{-1}$, to test for a trend. It is possible
that $\dot{M}_{out}$ is indeed very low for NGC~4395, but this value does not
include contributions from the X-ray absorbers (see the Appendix).

In Figure 6, we plot log($L_{KE}/L_{bol}$) against $L_{bol}$. Out of the six
moderate-luminosity AGN (i.e., excluding NGC~4395), three (NGC~4051, NGC~3516,
and NGC~3783) have kinetic luminosities that are approximately 0.5\% to 5\% of
their bolometric luminosities, which is the range typically assumed by feedback
models (Hopkins \& Elvis 2010, and references therein). NGC~5548 could
potentially be in this range, whereas NGC~4151 and NGC~7469 are at the
$\ltsim$0.1\% level. Once again, $L_{KE}/L_{bol}$ appears to be low for the
dwarf Seyfert 1 galaxy NGC~4395, but the values for the X-ray absorbers need to
be included to test this possibility. Excluding NGC~4395, we see no clear trend
in $\dot{M}_{out}$/$\dot{M}_{acc}$ or $L_{KE}/L_{bol}$ with either
$L_{bol}/L_{Edd}$ or black hole mass. However, the current sample is small, and
detailed studies of more AGN outflows are needed to explore the dependence of
feedback on fundamental AGN properties.

\section{Conclusions and Discussion}

The total mass outflow rates from UV and X-ray absorbers in the
moderate-luminosity AGN in our sample are typically 10 -- 1000 times the
accretion rates needed to provide the observed luminosities. The majority of
this outflow must therefore originate from outside the inner accretion disk.
There are two interesting possibilities. One is that a large reservoir of gas
has accumulated, in a torus or other circumnuclear structure, and the gas is
being accelerated off this structure (e.g., Krolik \& Kriss 2001). A related
possibility is that the gas is continuously accelerated directly off the fueling
flow over a range of distances. There is some evidence for the latter in larger
scale NLR outflows (Crenshaw et al. 2010; Fischer et al. 2010).

Previous AGN feedback models have typically required that $\sim$5\% of the
bolometric luminosity of an AGN be converted into kinetic luminosity in order to
regulate the growth of a SMBH and its galactic bulge (Di Matteo et al. 2005;
Hopkins et al. 2005). However, Hopkins \& Elvis (2010) have presented a model in
which only $\sim$0.5\% conversion is required. We find that the total kinetic
luminosity, summed over all absorbers, is 0.5\% to 5\% of the bolometric
luminosity for half of our moderate-luminosity AGN, in the range required by the
models. Of the remaining three, 2 of these have $L_{KE} \gtsim 0.1\% L_{bol}$
and one has $L_{KE} \ltsim 0.1\% L_{bol}$. However, we emphasize that some
absorbers still have no usable limits on $r$, so that future work may actually
increase these values of $\dot{M}_{out}$ and $L_{KE}$. NLR outflows, which we
consider in a subsequent paper, will further increase these values. Thus, we
find that the outflowing UV and X-ray absorbers in moderate-luminosity ($10^{43}
- 10^{45}$ ergs sec$^{-1}$) AGN have the potential to deliver significant
feedback to their environments.

\acknowledgments

This research has made use of the NASA/IPAC Extragalactic Database (NED)
which is operated by the Jet Propulsion Laboratory, California Institute of
Technology, under contract with the National Aeronautics and Space
Administration. This research has made use of NASA's Astrophysics Data
System Bibliographic Services.

\newpage
\appendix
\section{Details on Outflowing Absorbers in Individual AGN}

In Table 3, we give detailed measurements of individual absorption components
from the literature and the corresponding references in the subsection for
each AGN . The absorption component names are from the original studies. We
give $v_r$ for each UV and X-ray component, and the FWHM for each UV component
only. In cases where FWHM values were available for multiple lines, we chose
those values that corresponded to less saturated lines in order to minimize the
effects of saturation on broadening the observed profiles. The X-ray components
tend to be unresolved or barely resolved in {\it CXO} or {\it XMM-Newton}
grating spectra (Kaspi et al. 2003), and even the resolved profiles are often
likely blends of several distinct kinematic components (Gabel et al. 2003).
Furthermore, many of the quoted values are from the velocity dispersion used in
the photoionization model to fit the spectrum, and not from direct measurements.
Thus, we do not include FWHM values for the X-ray absorbers in our compilation.

For each component in Table 3 with an available photoionization model, we give
log($U$) and log(N$_H$). For studies where the ionization parameter $\xi$ was
given, we converted to log($U$) as described in Section 2. If there were
different values at different epochs due to variability, we averaged those
values to get the log($U$) and log($N_H$) in the table. For each UV absorber, we
give the size of the C IV BLR $r_{CIV}$ from Table 1 as the absolute minimum
values for $r$. We also give the absolute maximum value for $r$, $r_{>\Delta
r}$. In most cases, however, we do not use this value, because $r_{>\Delta r}$
is on the order of kpcs or more. However, it is useful for a few X-ray
absorbers.

The most crucial parameters for this study are our adopted $r_{min}$ and
$r_{max}$ in Table 3, which are the lower and upper limits for the radial
locations of the absorbers. In most cases, we were able to use values from the
literature based on variability of the absorption (or lack thereof) and/or
absorption (or its absence) from excited levels to provide density limits,
which, combined with $U$, yield limits to the radial locations. In a few
specific cases (e.g., in NGC~3783 and NGC~4151), the actual values have been
determined, and these are listed as both $r_{min}$ and $r_{max}$. When these
values are not available, we resort to $r_{CIV}$ and $r_{>\Delta r}$ when
appropriate.

Finally, we give minimum and maximum values for the feedback parameters
$\dot{M}_{out}$ and $L_{KE}$ in Table 3, based on the radial location limits.
We then sum the contributions from the UV and X-ray absorbers to obtain limits
for the total $\dot{M}_{out}$ and $L_{KE}$ in each AGN for use in Table 2. In a
few cases, particular UV and X-ray absorbers have been shown to arise in the
same gas, and we do not count both when determining the totals. If a particular
component has no maximum value, we use its minimum value when determining a
value for the total maximum $\dot{M}_{out}$ and $L_{KE}$. We give a detailed
discussion of these determinations for each AGN in the following subsections.

\subsection{NGC 3516}

NGC~3516 shows eight kinematic components of intrinsic absorption in the UV
(Component 1 has two physical subcomponents, 1a and 1b, with different $U$ and
$N_H$; Kraemer et al. 2002) and at least three distinct components in X-rays
(Turner et al. 2005, 2008, 2011). Based on detailed photoionization models and
shielding of the ionizing continuum by these components, Kraemer et al. showed
that the UV components follow the sequence 3+4 (blended at most epochs), 2, 1,
and 5 -- 8 in order of increasing radial distance from the central SMBH, and
Components 1 -- 4 are responsible for the X-ray absorption identified by Netzer
et al. (2002). We therefore do {\it not} use this X-ray component, labeled
``UV'' in Turner et al. (2005), in determining the total mass outflow rate and
kinetic luminosity. Lower limits to the radial distances of Components 1 -- 4
were determined from the lack of metastable C~III absorption by Kraemer et al.
(2002). As shown in Table 3, we have increased these limits to account for the
lack of C~III absorption at levels populated by lower densities than originally
considered (specifically, the transition from J $=$ 0; see Gabel et al. 2005).
The lower limit for Component 1 can also be applied to UV components 5 -- 8, but
no reasonable upper limits are available for these components. An upper limit to
the radial location of ``UV'' of $\sim$0.4 pc was determined from the
variability
of O~VII absorption in the X-rays (Netzer et al. 2002; Kraemer et al. 2002).
This upper limit is smaller than the lower limits for UV components 1 and 2, but
not 3+4, suggesting that it was the latter UV component that varied (consistent
with its previous strong variability; Kraemer et al. 2002).

The ionizing radiation for the other two X-ray components in Turner et al.
(2005), ``Hi'' and ``Heavy'', cannot be shielded by ``UV'' (Kraemer et al.
2002), and they must therefore lie inside of the latter, providing an upper
limit to their radial distances. The $\Delta r/r < 1$ requirement puts a tighter
constraint on X-ray component Heavy -- it must be lie at $r \leq 0.11$ pc. Heavy
shows partial covering in the line of sight and possibly higher columns than
listed in Table 3 at some epochs (Turner et al. 2008; 2011), so its upper limits
for mass outflow rates and kinetic luminosities are approximate. Upper limits
are not available for most of the UV components, so the maximum values for
$\dot{M}_{out}$ and $L_{KE}$ in NGC~3516 could be higher.

\subsection{NGC 3783}

NGC~3783 has three distinct kinematic components of intrinsic UV absorption
(Kraemer et al. 2001; Gabel et al. 2005) and three components of X-ray
absorption (Netzer et al. 2003).  UV component 1 consists of two physical
subcomponents (1a and 1b); we assume these are co-located because they are at
the same velocity. Metastable C~III absorption provides the density and distance
of UV 1b. Monitoring of the strong absorption variations in UV components 2 and
3 provides upper limits to their distances (Gabel et al. 2005) and absorption of
the BLR provides lower limits. X-ray component XLI has similar $U$ and $v_r$
coverage as components 1b, 2, and 3, suggesting it arises in the same gas, but
it has $\sim$3 times the N$_H$ of the UV absorbers (Gabel et al. 2005). We
therefore exclude XLI in our lower limits, but include its remaining column,
after subtracting the UV columns in our upper limits, for outflow rates and
kinetic luminosities. The lack of variability in the X-ray components on a time
scale of $\sim$10 days, despite strong continuum variations, provides lower
limits to the radial locations of X-ray components XMI and XHI (Netzer et al.
2003). We use the $\Delta$r/r constraints for upper limits to the radial
locations of XMI and XHI, and set the upper limit of XLI to that of the most
distant UV component (3), as discussed above. Krongold et al. (2005) find an
upper limit of $\sim$6 pc for this lower-ionization component, which is
consistent with the other determinations.

\subsection{NGC 4051}

NGC~4051 shows 9 distinct components of UV absorption in its STIS (Collinge et
al. 2001) and COS (Kraemer et al. 2012) spectra, in addition to a component that
is clearly due to the ISM in our Galaxy. Component 1 likely arises in our Galaxy
and Components 8 and 9 (plus Collinge et al.'s Component 10) likely arise in the
host galaxy of NGC~4051. Components 3, 4, and 6 are weak and difficult to
separate from the stronger Components 2, 5, and 7. The latter provide the bulk
of the outflow in the UV and are modeled by Kraemer et al. (2012). Steenbrugge
et al. (2009) use 4 zones to model the X-ray absorption, similar to those in
Lobban et al. (2011). We use Steenbrugge et al.'s components because these
authors provide distance constraints. Based on photoionization parameters and
velocity correspondences, Kraemer et al. (2012) find that Steenbrugge et al.'s
X-ray 1 and 2 are likely the same as UV 7 and 5, respectively, and thus we do
not include these UV components in our totals for mass outflow rates and kinetic
luminosities. Steenbrugge et al.'s X-ray 3 is at the same approximate velocity
as UV 2, suggesting that they are co-located, but their $U$ and $N_H$ are very
different. We therefore include both in our calculations. Based on the lack of
variability in X-ray 1, 3, and 4, Steenbrugge et al. obtained lower limits to
their locations. The $r > \Delta r$ requirement for Component 4 gives an upper
limit consistent with the above lower limit to within the uncertainty of
$\sim$0.3 dex.Component 2 varied over this time period, providing an upper
limit to its radial location.

Steenbrugge et al.'s Component 4 has unusually high $v_r$, $U$, and $N_H$, and
it dominates the mass outflow parameters. However, it does not qualify as an
ultrafast outflow (Tombesi et al. 2010). Thus, the global covering factor of
this component is unclear. We therefore use $C_g = 0.5$ for the upper limits to
its $\dot{M}_{out}$ and $L_{KE}$, but scale these numbers by a factor of 0.1
(i.e., $C_g = 0.05$) for the lower limits to account for its uniqueness among
our sample of 10 Seyfert 1 galaxies.

\subsection{NGC 4151}

NGC~4151 has many components of UV absorption that were first identified by
Weymann et al. (1997). Their D and E components are not separable and are lumped
together as a single kinematic component ``D+E'' in subsequent studies (e.g.,
Crenshaw et al. 2000; Kraemer et al. 2001). Components D+E and E$'$ are modeled
as four (a, b, c, d) and two (a, b) physical subcomponents, respectively, in
Kraemer et al. (2005, 2006), who found that subcomponent D+Ea is responsible for
the low-ionization X-ray absorption. Another component (``Xhigh'') is needed to
explain the high-ionization absorption lines in the X-ray region of the
spectrum. Distances to all of the UV absorbers are obtained from absorption
lines arising from metastable and/or fine-structure excited levels. All D+E
subcomponents are assumed to lie at the same distance as D+Ea (0.1 pc) due to
their velocity correspondence. Component D$'$ is screened by D+Ea and must lie
outside of it. Xhigh has the same approximate radial velocity of D+Ea and must
lie inside of the latter, because photoionization models demonstrate that Xhigh
is not shielded from the ionizing radiation by D+Ea (Kraemer et al. 2005). UV
components A and C are in the NLR at large distances from the central SMBH, and
we include these only in the upper limits for mass outflow rates and kinetic
luminosities, even though their distances are known, because they may not have
global covering factors as large as 0.5.

\subsection{NGC 4395}
NGC~4395 is a nearby dwarf Seyfert 1 (Filippenko \& Sargent 1989), with a very
low luminosity (L$_{bol} \approx 5 \times 10^{40}$ ergs s$^{-1}$) and black-hole
mass (M $\approx 3.6 \times 10^{5}$ M$_{\odot}$) (Peterson et al. 2005).
Nevertheless, its UV spectrum shows two UV absorbers that are outflowing from
its nucleus (Crenshaw et al. 2004, and references therein). Baskin \& Laor
(2008) determined the physical conditions in the two absorbers and pointed out
that they are likely between its tiny C~IV BLR ($\sim$ 0.4 light days, Peterson
et al. 2004) and inner NLR ($\sim2.3 \times$ 10$^{-2}$ pc, Kraemer et al. 1999).
Moreover, NGC 4395 shows strong evidence for warm absorbers in its X-ray spectra
(Iwasawa et al. 2000; Shih et al. 2003; Moran et al. 2005). We give Shih et
al.'s characterization of a constant and a variable zone of ionized absorption
in the X-rays. Unfortunately, we are unable to include feedback parameters for
the X-ray absorbers, because there has been no determination of their radial
velocities.

\subsection{Mrk 279}

Mrk 279 has five principal components of UV absorption; Component 1 likely
arises in the host galaxy and the low ionization lines of Component 4 may arise
in its halo or a companion galaxy (Scott et al. 2004, 2009; Gabel et al. 2005).
Arav et al. (2007) used photoionization models to determine the physical
conditions in Component 2. Scott et al. (2009) find that variability of the UV
absorption lines is due to both varying contributions from emission regions that
have different covering factors and intrinsic variation, and the locations of
the absorbers have not been determined. Ebrero et al. (2010) find two warm
absorbers in X-ray spectra (see also Constantini et al. 2007). X-ray component 1
and UV 2 have similar $v_r$ and $U$, but the former has $\sim$10 times higher
$N_H$. Ebrero et al. find no strong evidence for or against variability in the
X-ray absorbers, and thus no distances or feedback parameters are available for
this AGN.

\subsection{NGC 5548}

NGC~5548 shows five principal components of intrinsic UV absorption (Crenshaw et
al. 1999, 2003b; Mathur et al. 1999). None of the components show evidence for
strong variability in $U$ despite large-scale continuum changes, indicating
distances $>$ 70 pc from the central SMBH (Crenshaw et al. 2009). However, a
large portion of Component 3 is responsible for some of the X-ray absorption
described in Steenbrugge et al. (2005), which Detmers et al. (2008) place at a
distance $<$ 7 pc from the SMBH, based on ionization changes.
Andrade-Vel{\'a}zquez et al. (2010) find four components of X-ray absorption:
high-velocity super-high ionization  phase (HV-SHIP), high-velocity
high-ionization phase (HV-HIP), low-velocity high-ionization phase (LV-HIP), and
low-velocity low-ionization phase (LV-LIP). Krongold et al. (2010) find no
response of HV-SHIP to changes in the ionizing continuum, and we assume that
HV-HIP is co-located, putting the HV components at distances $>$ 0.03 pc from
the SMBH. The two LV components are very similar to UV Component 3 in radial
velocity, average $U$, and total $N_H$, so we do not use these for the total
$\dot{M}_{out}$ and $L_{KE}$. Krongold et al. find a possible response of the
LV-LIP component, and assuming LV-HIP is co-located, these components are at
distances $<$ 3 pc, similar to the limit found by Detmers et al (2008). We have
no upper limits for components other than UV 3 (LV-HIP+LV-LIP), and we therefore
do not include upper limits for total $\dot{M}_{out}$ and $L_{KE}$.

\subsection{Mrk 509}

Mrk~509 has at least seven components of UV absorption (Kriss et al. 2000;
Kraemer et al. 2003; Kriss et al. 2011), depending on how the absorption
structure is subdivided. Kraemer et al. (2003) give $U$ and $N_H$ for the UV
components. Kriss et al. (2011) find that Components 1 -- 3 arise in outflows,
whereas Components 4 -- 7 are close to systemic or at positive velocities,
indicating origins in the host galaxy, halo, or other regions not associated
with the outflows. Kriss et al. (2011) find variability in UV 1, indicating a
radial location $<$ 250 pc from the central SMBH. Although the UV and X-ray
absorbers have the same radial velocity coverage, the latter have higher $U$ and
$N_H$ (Kriss et al. 2011). Ebrero et al. (2011) give the physical conditions for
the three X-ray absorbers in Mrk~509, which are similar to those found by
Detmers et al. (2010, 2011). Detmers et al. use mild variability in the highest
ionization component (X-ray 3 in the table) to place it at a distance $<$ 0.5 pc
from the SMBH. Because we have distances for only one UV and one X-ray absorber,
we do not include Mrk~509 in our determination of total $\dot{M}_{out}$ and
$L_{KE}$.

\subsection{Akn 564}

Akn~564 shows strong UV absorption lines from a ``lukewarm'' absorber that
reddens the NLR in this NLS1 (Crenshaw et al. 2002), placing it at a distance
$>$ 95 pc from the central SMBH. The absorber has a radial velocity of $-$190 km
s$^{-1}$, resulting in significant lower limits to $\dot{M}_{out}$ and $L_{KE}$,
assuming a global covering factor of $C_g = 0.5$. However, we find that C$_g
\leq 0.05$ for this absorber; otherwise its emission-lines fluxes for the higher
ionization lines would exceed those observed in the NLR (Crenshaw et al. 2002).
Thus, the lower limits for $\dot{M}_{out}$ and $L_{KE}$ should probably be
divided by 10. Smith et al. (2008) find 3 warm absorbers in X-ray spectra of
Akn~564, and two of these could produce at least some of the UV absorption
(Matsumoto et al. 2004) . There are no reliable distances for the X-ray
absorbers, and, given the above concerns about the UV absorber, we do not
include this AGN in our determinations of $\dot{M}_{out}$ and $L_{KE}$.

\subsection{NGC 7469}

The UV spectrum of NGC~7469 has two main kinematic components (Kriss et al.
2000, 2003) that have been modeled by Scott et al. (2005). Scott et al. detect
variability of the H~I column density in both components, yielding upper limits
to their distances. Blustin et al. (2007) give details on three X-ray warm
absorbers in NGC~7469. X-ray component 1 is close to UV 2 in both $v_r$ and $U$,
but has a higher $N_H$; nevertheless, the photoionization model for X-ray 1
predicts UV ionic columns that are reasonably good matches to the observed
values (Blustin et al. 2007). X-ray components 2 and 3 are similar to UV 1 in
$v_r$, but have much higher $U$ and $N_H$. There is no information on
variability of the X-rays absorbers, but reasonable upper limits can be derived
for components 2 and 3 from the $\Delta$r/r constraint. We include the resulting
upper limits for $\dot{M}_{out}$ and $L_{KE}$ in our overall feedback
determinations.

\clearpage

\newpage

\begin{deluxetable}{lrrccccc}
\tablecolumns{8}
\footnotesize
\tablecaption{AGN Fundamental Parameters}
\tablewidth{0pt}
\tablehead{
\colhead{Name} & \colhead{$r_{H\beta}$} & \colhead {$r_{CIV}$} &
\colhead{log ($L_{bol}$)} & \colhead{log ($M_{BH}$)} &
\colhead{$L_{bol}/L_{Edd}$} & \colhead{($\dot{M}_{acc}$)} &
\colhead{Reference}\\
\colhead{} & \colhead{(ltday)} & \colhead{(ltday)} & \colhead{(ergs s$^{-1}$)} &
\colhead{(M$_{\odot}$)} &
\colhead{} & \colhead{(M$_{\odot}$ yr$^{-1}$)} & \colhead{}
}
\startdata
NGC 3516	&11.7	&	&44.16	&7.50	&0.036	&0.026 & 1 \\
NGC 3783	&10.2	&3.8	&44.25	&7.47	&0.047	&0.032 & 2 \\
NGC 4051	&1.9	&	&42.81	&6.24	&0.030	&0.001 & 1 \\
NGC 4151	&6.6	&	&43.87	&7.66	&0.013	&0.013 & 3 \\
NGC 4395	&	&0.04	&40.77	&5.56	&0.001	&1.06E-5 & 4 \\
Mrk 279		&12.5	&	&44.87	&7.54	&0.169	&0.134 & 2 \\
NGC 5548	&12.4	&8.3	&43.90	&7.64	&0.014	&0.014 & 1 \\
Mrk 509		&79.6	&	&45.27	&8.16	&0.104	&0.336 & 2 \\
Akn 564		&17.9	&	&44.61	&6.42	&1.242	&0.074 & 5 \\
NGC 7469	&4.5	&2.5	&44.71	&7.09	&0.335	&0.093 & 2 \\
\enddata
\tablerefs{(1) Denney et al. 2010, (2) Peterson et al. 2004, (3) Bentz et al.
2006; Peterson et al. 2004, (4) Peterson et al. 2005, 2006, (5) Botte et al.
2004. For $r_{CIV}$, the references are Peterson et al. (2004, 2005). }

\end{deluxetable}

\begin{deluxetable}{lcccc}
\tablecolumns{5}
\footnotesize
\tablecaption{AGN Feedback Parameters}
\tablewidth{0pt}
\tablehead{
\colhead{Name} & \colhead{$\dot{M}_{out}$} & \colhead{log (L$_{KE}$)} &
\colhead{$\dot{M}_{out}$/$\dot{M}_{acc}$} & \colhead{L$_{KE}$/L$_{bol}$} \\
\colhead{} & \colhead{(M$_{\odot}$ yr$^{-1}$)} & \colhead{(ergs s$^{-1}$)} &
\colhead{} & \colhead{}
}
\startdata
NGC 3516 &3.8E+00 -- 7.7E+00 &41.73 -- 42.54 &150 -- 300 &3.7E-03 -- 2.4E-02 \\
NGC 3783 &5.2E+00 -- 3.0E+01 &42.45 -- 42.88 &160 -- 940 &1.6E-02 -- 4.3E-02\\
NGC 4051 &1.4E-02 -- 1.3E-01 &40.93 -- 41.93 &12 -- 110  &1.3E-02 -- 1.3E-01\\
NGC 4151 &3.2E-01 -- 6.7E-01 &40.40 -- 41.18 &24 -- 50   &3.4E-04 -- 2.0E-03\\	
NGC 4395 &2.0E-07 -- 1.4E-04 &33.91 -- 36.74 &0.02 -- 13  &1.4E-07 -- 9.4E-05\\
NGC 5548 &$>$7.9E-01   &$>$41.26	&$>$55	&$>$2.3E-03 \\
NGC 7469 &$<$5.7E+00   &$<$42.07 &$<$61  &$<$2.3E-03 \\	
\enddata
\end{deluxetable}

\pagestyle{empty}
\tabletypesize{\footnotesize}
\topmargin 1.0in
\oddsidemargin -0.5in
\textwidth 7.5in

\setcounter{table}{2}
\begin{deluxetable}{lcccccccccccccc} 
\tablecolumns{13} 
\rotate
\tablewidth{0pc} 
\tablecaption{Feedback Parameters for Individual Kinematic Components} 
\tablehead{ 
\colhead{Comp.} & \colhead{$v_r$} & \colhead{FWHM} & \colhead{log(U)} &
\colhead{log(N$_H$)} & \colhead{log($r_{CIV}$)} &
\colhead{log($r_{>\Delta r}$)} & \colhead{log(r$_{min}$)} &
\colhead{log(r$_{max}$)} &
\colhead{$\dot{M}_{out~min}$} & \colhead{$\dot{M}_{out~max}$} &
\colhead{$L_{KE~min}$} & \colhead{$L_{KE~max}$}\\
\colhead{} & \colhead{(km s$^{-1}$)} & \colhead{(km s$^{-1}$)} & \colhead{} &
\colhead{(cm$^{-2}$)} & \colhead{(cm)} &
\colhead{(cm)} & \colhead{(cm)} &\colhead{(cm)} &
\colhead{(M$_{\odot}$ yr$^{-1}$)} & \colhead{(M$_{\odot}$ yr$^{-1}$)} & 
\colhead{(ergs s$^{-1}$)} & \colhead{(ergs s$^{-1}$)}
}
\startdata 
\hline 
\multicolumn{13}{c}{NGC 3516 UV Absorption} \\
\hline 
     1a      & -376 &   70 &  -0.34 &  21.38 &  16.00 &  21.06 &  19.68 &    &
1.01E+00 &  & 4.48E+40 &  \\
     1b      & -376 &   70 &  -2.35 &  18.85 &  16.00 &  25.60 &  19.68 &    &
2.95E-03 &  & 1.31E+38 &  \\
      2      & -183 &   44 &  -0.39 &  21.59 &  16.00 &  20.90 &  19.28 &    &
3.18E-01 &  & 3.34E+39 &  \\
     3+4     &  -36 &   31 &  -0.29 &  21.15 &  16.00 &  21.24 &  17.19 &  18.08
& 1.83E-04 & 1.42E-03 & 7.43E+34 & 5.76E+35 \\
      5      &-1372 &  271 &   0.81 &  19.18 &  16.00 &  22.11 &  19.68 &    &
2.30E-02 &  & 1.36E+40 &  \\
      6      & -994 &   36 &   0.93 &  20.18 &  16.00 &  20.99 &  19.68 &    &
1.67E-01 &  & 5.18E+40 &  \\
      7      & -837 &   99 &   1.03 &  21.30 &  16.00 &  19.77 &  19.68 &    &
1.87E+00 &  & 4.12E+41 &  \\
      8      & -692 &   35 &   1.03 &  20.74 &  16.00 &  20.33 &  19.68 &    &
4.26E-01 &  & 6.40E+40 &  \\
  UV 1 - 8   &      &      &    &    &    &    &    &    & 3.82E+00 & 3.82E+00 &
5.41E+41 & 5.90E+41 \\
\hline 
\multicolumn{13}{c}{NGC 3516 X-ray Absorption} \\
\hline 
     UV      & -300 &      &  -0.90 &  21.70 &    &  21.30 &  17.19 &  18.08 &
5.45E-03 & 4.23E-02 & 1.54E+38 & 1.19E+39 \\
     Hi      &-1140 &      &   1.60 &  22.20 &    &  18.30 &    &  18.08 &  &
5.08E-01 &  & 2.07E+41 \\
    Heavy    &-1600 &      &   1.15 &  23.40 &    &  17.55 &    &  17.55 &  &
3.32E+00 &  & 2.67E+42 \\
  Hi+Heavy   &      &      &    &    &    &    &    &    &  & 3.83E+00 &  &
2.87E+42 \\
  UV+X-ray   &      &      &    &    &    &    &    &    & 3.82E+00 & 7.65E+00 &
5.41E+41 & 3.46E+42 \\
\hline 
\multicolumn{13}{c}{NGC 3783 UV Absorption} \\
\hline 
     1a      &-1365 &  193 &  -1.60 &  20.60 &  15.99 &  23.19 &  19.94 &  19.94
& 1.11E+00 & 1.11E+00 & 6.47E+41 & 6.47E+41 \\
     1b      &-1365 &  193 &  -0.40 &  21.10 &  15.99 &  21.49 &  19.94 &  19.94
& 3.50E+00 & 3.50E+00 & 2.05E+42 & 2.05E+42 \\
      2      & -548 &  170 &  -0.45 &  20.40 &  15.99 &  22.24 &  15.99 &  19.90
& 3.17E-05 & 2.56E-01 & 2.99E+36 & 2.41E+40 \\
      3      & -724 &  280 &  -0.50 &  21.10 &  15.99 &  21.59 &  15.99 &  20.20
& 2.10E-04 & 3.38E+00 & 3.95E+37 & 5.56E+41 \\
   UV 1-3    &      &      &    &    &    &    &    &    & 4.61E+00 & 8.25E+00 &
2.69E+42 & 3.27E+42 \\
\hline 
\multicolumn{13}{c}{NGC 3783 X-ray Absorption} \\
\hline 
     XLI     & -800 &      &  -0.50 &  21.90 &    &  20.79 &  19.00 &  20.20 &
1.49E+00 & 1.54E+01 & 2.99E+41 & 3.08E+42 \\
     XMI     & -800 &      &   0.80 &  22.00 &    &  19.39 &  18.30 &  19.39 &
3.74E-01 & 4.57E+00 & 7.50E+40 & 9.18E+41 \\
     XHI     & -800 &      &   1.30 &  22.30 &    &  18.59 &  17.70 &  18.59 &
1.87E-01 & 1.45E+00 & 3.76E+40 & 2.90E+41 \\
   XMI+XHI   &      &      &    &    &    &    &    &    & 5.61E-01 & 2.14E+01 &
1.13E+41 & 4.29E+42 \\
  UV+X-ray   &      &      &    &    &    &    &    &    & 5.20E+00 & 2.96E+01 &
2.81E+42 & 7.56E+42 \\
\hline 
\multicolumn{13}{c}{NGC 4051 UV Absorption} \\
\hline 
      1      & -647 &   40 &    &    &  15.21 &    &    &    &  &  &  &  \\
      2      & -505 &  165 &  -0.72 &  20.17 &  15.21 &  21.30 &  16.70 &    &
8.76E-05 &  & 7.01E+36 &  \\
      3      & -430 &   63 &    &    &  15.21 &    &    &    &  &  &  &  \\
      4      & -337 &   52 &    &    &  15.21 &    &    &    &  &  &  &  \\
      5      & -268 &  133 &  -0.68 &  20.34 &  15.21 &  21.09 &    &  17.48 & 
& 4.14E-04 &  & 9.33E+36 \\
      6      & -158 &   45 &    &    &  15.21 &    &    &    &  &  &  &  \\
      7      & -107 &   64 &  -0.80 &  20.18 &  15.21 &  21.37 &  17.95 &    &
3.38E-04 &  & 1.21E+36 &  \\
      8      &  -48 &   84 &    &    &  15.21 &    &    &    &  &  &  &  \\
      9      &   30 &   23 &    &    &  15.21 &    &    &    &  &  &  &  \\
    UV 2     &      &      &    &    &    &    &    &    & 8.76E-05 & 8.76E-05 &
7.01E+36 & 7.01E+36 \\
\hline 
\multicolumn{13}{c}{NGC 4051 X-ray Absorption} \\
\hline 
      1      & -210 &      &  -1.43 &  20.08 &    &  22.10 &  17.95 &    &
5.27E-04 &  & 7.28E+36 &  \\
      2      & -200 &      &  -0.63 &  20.46 &    &  20.92 &    &  17.48 &  &
4.08E-04 &  & 5.11E+36 \\
      3      & -580 &      &   0.82 &  20.90 &    &  19.03 &  16.70 &    &
5.40E-04 &  & 5.70E+37 &  \\
      4      &-4670 &      &   1.69 &  22.30 &    &  16.76 &  16.85 &  16.76 &
1.25E-02 & 1.25E-01 & 8.54E+40 & 8.54E+41 \\
    X-ray    &      &      &    &    &    &    &    &    & 1.35E-02 & 1.26E-01 &
8.55E+40 & 8.54E+41 \\
  UV+X-ray   &      &      &    &    &    &    &    &    & 1.36E-02 & 1.26E-01 &
9.55E+40 & 8.54E+41 \\
\hline 
\multicolumn{13}{c}{NGC 4151 UV Absorption} \\
\hline 
      A      &-1588 &   36 &  -2.92 &  18.10 &  15.76 &  26.63 &  21.32 &  21.32
& 9.77E-02 & 9.77E-02 & 7.73E+40 & 7.73E+40 \\
      C      & -858 &   27 &  -2.92 &  18.00 &  15.76 &  26.73 &  21.81 &  21.81
& 1.30E-01 & 1.30E-01 & 2.99E+40 & 2.99E+40 \\
    D+Ea     & -491 &  435 &  -0.39 &  22.93 &  15.76 &  19.27 &  17.49 &  17.49
& 3.02E-01 & 3.02E-01 & 2.29E+40 & 2.29E+40 \\
    D+Eb     & -491 &  435 &  -1.67 &  20.80 &  15.76 &  22.68 &  17.49 &  17.49
& 2.24E-03 & 2.24E-03 & 1.69E+38 & 1.69E+38 \\
    D+Ec     & -491 &  435 &  -1.08 &  21.60 &  15.76 &  21.29 &  17.49 &  17.49
& 1.41E-02 & 1.41E-02 & 1.07E+39 & 1.07E+39 \\
    D+Ed     & -491 &  435 &  -3.35 &  19.50 &  15.76 &  25.66 &  17.49 &  17.49
& 1.12E-04 & 1.12E-04 & 8.49E+36 & 8.49E+36 \\
     D'      &-1680 &  940 &   0.31 &  20.00 &  15.76 &  21.50 &  17.49 &  18.49
& 1.21E-03 & 1.21E-02 & 1.08E+39 & 1.08E+40 \\
     E'a     & -215 &   59 &  -1.74 &  20.60 &  15.76 &  22.95 &  18.27 &  18.27
& 3.73E-03 & 3.73E-03 & 5.41E+37 & 5.41E+37 \\
     E'b     & -215 &   59 &  -3.64 &  19.00 &  15.76 &  26.45 &  18.27 &  18.27
& 9.37E-05 & 9.37E-05 & 1.36E+36 & 1.36E+36 \\
     UV      &      &      &    &    &    &    &    &    & 3.24E-01 & 5.62E-01 &
2.52E+40 & 1.42E+41 \\
\hline 
\multicolumn{13}{c}{NGC 4151 X-ray Absorption} \\
\hline 
    Xhigh    & -491 &      &   1.05 &  22.50 &    &  18.26 &    &  17.49 &  &
1.12E-01 &  & 8.49E+39 \\
  UV+X-ray   &      &      &    &    &    &    &    &    & 3.20E-01 & 6.70E-01 &
2.52E+40 & 1.51E+41 \\
\hline 
\multicolumn{13}{c}{NGC 4395 UV Absorption} \\
\hline 
    1 (B)    & -840 &      &  -1.70 &  19.00 &  14.02 &  21.41 &  14.02 &  16.85
& 2.04E-08 & 1.39E-05 & 4.51E+33 & 3.08E+36 \\
   2 (Ah)    & -250 &      &  -0.70 &  20.48 &  14.02 &  18.93 &  14.02 &  16.85
& 1.83E-07 & 1.25E-04 & 3.59E+33 & 2.45E+36 \\
     UV      &      &      &    &    &    &    &    &    & 2.03E-07 & 1.39E-04 &
8.10E+33 & 5.53E+36 \\
\hline 
\multicolumn{13}{c}{NGC 4395 X-ray Absorption} \\
\hline 
  Constant   &      &      &   0.80 &  22.39 &    &  15.52 &    &  15.52 &
&  &  &  \\
  Variable   &      &      &   1.18 &  22.90 &    &  14.63 &    &  14.63 &
&  &  &  \\
    X-ray    &      &      &    &    &    &    &    &    &  &  &  &  \\
  UV+X-ray   &      &      &    &    &    &    &    &    &  &  &  &  \\
\hline 
\multicolumn{13}{c}{Mrk 279 UV Absorption} \\
\hline 
      1      &   90 &      &    &    &    &    &    &    &  &  &  &  \\
      2      & -265 &      &  -1.00 &  18.90 &  16.03 &  24.91 &    &    &  &  &
 &  \\
      3      & -385 &      &    &    &    &    &    &    &  &  &  &  \\
      4      & -450 &      &    &    &    &    &    &    &  &  &  &  \\
      5      & -540 &      &    &    &    &    &    &    &  &  &  &  \\
     UV      &      &      &    &    &    &    &    &    &  &  &  &  \\
\hline 
\multicolumn{13}{c}{MRK 279 X-ray Absorption} \\
\hline 
      1      & -200 &      &  -0.80 &  19.85 &    &  23.76 &    &    &  &  &  & 
\\
      2      & -370 &      &   1.10 &  20.43 &    &  21.28 &    &    &  &  &  & 
\\
    X-ray    &      &      &    &    &    &    &    &    &  &  &  &  \\
\hline 
\multicolumn{13}{c}{NGC 5548 UV Absorption} \\
\hline 
      1      &-1040 &  220 &  -0.59 &  19.96 &  16.33 &  22.47 &  20.33 &    &
4.75E-01 &  & 1.61E+41 &  \\
      2      & -670 &   40 &  -0.78 &  19.20 &  16.33 &  23.42 &  20.33 &    &
5.31E-02 &  & 7.48E+39 &  \\
      3      & -530 &  160 &  -0.15 &  21.77 &  16.33 &  20.22 &    &  19.00 & 
& 7.30E-01 &  & 6.43E+40 \\
      4      & -340 &  150 &  -0.67 &  20.13 &  16.33 &  22.38 &  20.33 &    &
2.29E-01 &  & 8.32E+39 &  \\
      5      & -170 &   60 &  -0.67 &  19.41 &  16.33 &  23.10 &  20.33 &    &
2.19E-02 &  & 1.98E+38 &  \\
   UV 1-5    &      &      &    &    &    &    &    &    & 7.79E-01 &  &
1.77E+41 &  \\
\hline 
\multicolumn{13}{c}{NGC 5548 X-ray Absorption} \\
\hline 
   HV-SHIP   &-1040 &      &   1.23 &  21.73 &    &  18.88 &  17.00 &    &
1.31E-02 &  & 4.43E+39 &  \\
   HV-HIP    &-1180 &      &   0.67 &  21.03 &    &  20.14 &  17.00 &    &
2.96E-03 &  & 1.29E+39 &  \\
   LV-HIP    & -400 &      &   0.67 &  21.26 &    &  19.91 &    &  19.00 &  &
1.70E-01 &  & 8.55E+39 \\
   LV-LIP    & -590 &      &  -0.49 &  20.75 &    &  21.58 &    &  19.00 &  &
7.76E-02 &  & 8.48E+39 \\
    X-ray    &      &      &    &    &    &    &    &    & 1.60E-02 &  &
5.73E+39 &  \\
  UV+X-ray   &      &      &    &    &    &    &    &    & 7.95E-01 &  &
1.83E+41 &  \\
\hline 
\multicolumn{13}{c}{MRK 509 UV Absorption} \\
\hline 
      1      & -422 &   28 &  -0.82 &  19.01 &  16.84 &  25.02 &    &  20.89 & 
& 7.80E-02 &  & 4.35E+39 \\
      2      & -328 &   49 &  -1.31 &  18.92 &  16.84 &  25.60 &    &    &  &  &
 &  \\
      3      & -259 &   41 &  -1.48 &  18.25 &  16.84 &  26.44 &    &    &  &  &
 &  \\
     4'      &  -62 &   32 &  -1.19 &  18.53 &  16.84 &  25.87 &    &    &  &  &
 &  \\
     4h      &  -22 &   52 &  -1.02 &  18.97 &  16.84 &  25.26 &    &    &  &  &
 &  \\
     4l      &  -21 &   21 &  -1.70 &  19.53 &  16.84 &  25.38 &    &    &  &  &
 &  \\
      5      &   34 &   35 &  -0.82 &  18.59 &  16.84 &  25.44 &    &    &  &  &
 &  \\
      6      &  124 &   29 &  -0.78 &  18.79 &  16.84 &  25.20 &    &    &  &  &
 &  \\
      7      &  210 &   53 &  -0.16 &  19.60 &  16.84 &  23.77 &    &    &  &  &
 &  \\
     UV      &      &      &    &    &    &    &    &    &  & 7.80E-02 &  &
4.35E+39 \\
\hline 
\multicolumn{13}{c}{Mrk 509 X-ray Absorption} \\
\hline 
      1      &   73 &      &  -0.44 &  20.27 &    &  23.38 &    &    &  &  &  & 
\\
      2      & -196 &      &   0.76 &  20.73 &    &  21.72 &    &    &  &  &  & 
\\
      3      & -455 &      &   1.65 &  20.78 &    &  20.78 &    &  18.19 &  &
9.94E-03 &  & 6.45E+38 \\
    X-ray    &      &      &    &    &    &    &    &    &  & 9.94E-03 &  &
6.45E+38 \\
  UV+X-ray   &      &      &    &    &    &    &    &    &  & 8.79E-02 &  &
5.00E+39 \\
\hline 
\multicolumn{13}{c}{Akn 564 UV Absorption} \\
\hline 
      1      & -190 &  180 &  -1.48 &  21.21 &  16.19 &  22.82 &  20.47 &    &
2.11E+00 &  & 2.39E+40 &  \\
\hline 
\multicolumn{13}{c}{Akn564 X-ray Absorption} \\
\hline 
      1      &      &      &  -2.36 &  19.95 &    &  24.96 &    &    &  &  &  & 
\\
      2      &  -40 &      &  -0.63 &  20.38 &    &  22.80 &    &    &  &  &  & 
\\
      3      &  -10 &      &   1.06 &  20.78 &    &  20.71 &    &    &  &  &  & 
\\
    X-ray    &      &      &    &    &    &    &    &    &  &  &  &  \\
  UV+X-ray   &      &      &    &    &    &    &    &    &  &  &  &  \\
\hline 
\multicolumn{13}{c}{NGC 7469 UV Absorption} \\
\hline 
      1      & -560 &      &0.00    &  20.00 &  15.81 &  22.65 &    &  20.49 & 
& 4.04E-01 &  & 3.98E+40 \\
      2      &-1900 &      &  -1.10 &  18.60 &  15.81 &  25.15 &    &  21.27 & 
& 3.28E-01 &  & 3.71E+41 \\
 UV 1-2     &      &      &    &    &    &    &    &    &  & 7.32E-01 &  &
4.11E+41 \\
\hline 
\multicolumn{13}{c}{NGC 7469 X-ray Absorption} \\
\hline 
      1      &-2300 &      &  -0.70 &  19.48 &    &  23.87 &    &  23.87 &
&  &  &  \\
      2      & -720 &      &   1.23 &  21.30 &    &  20.12 &    &  20.12 &  &
4.41E+00 &  & 7.17E+41 \\
      3      & -580 &      &   2.06 &  21.46 &    &  19.13 &    &  19.13 &  &
5.25E-01 &  & 5.54E+40 \\
X-ray 2-3     &      &      &    &    &    &    &    &    &  & 4.93E+00 & 
&
7.72E+41 \\
  UV+X-ray   &      &      &    &    &    &    &    &    &  & 5.67E+00 &  &
1.18E+42 \\
\enddata 
\end{deluxetable} 

\clearpage

\figcaption[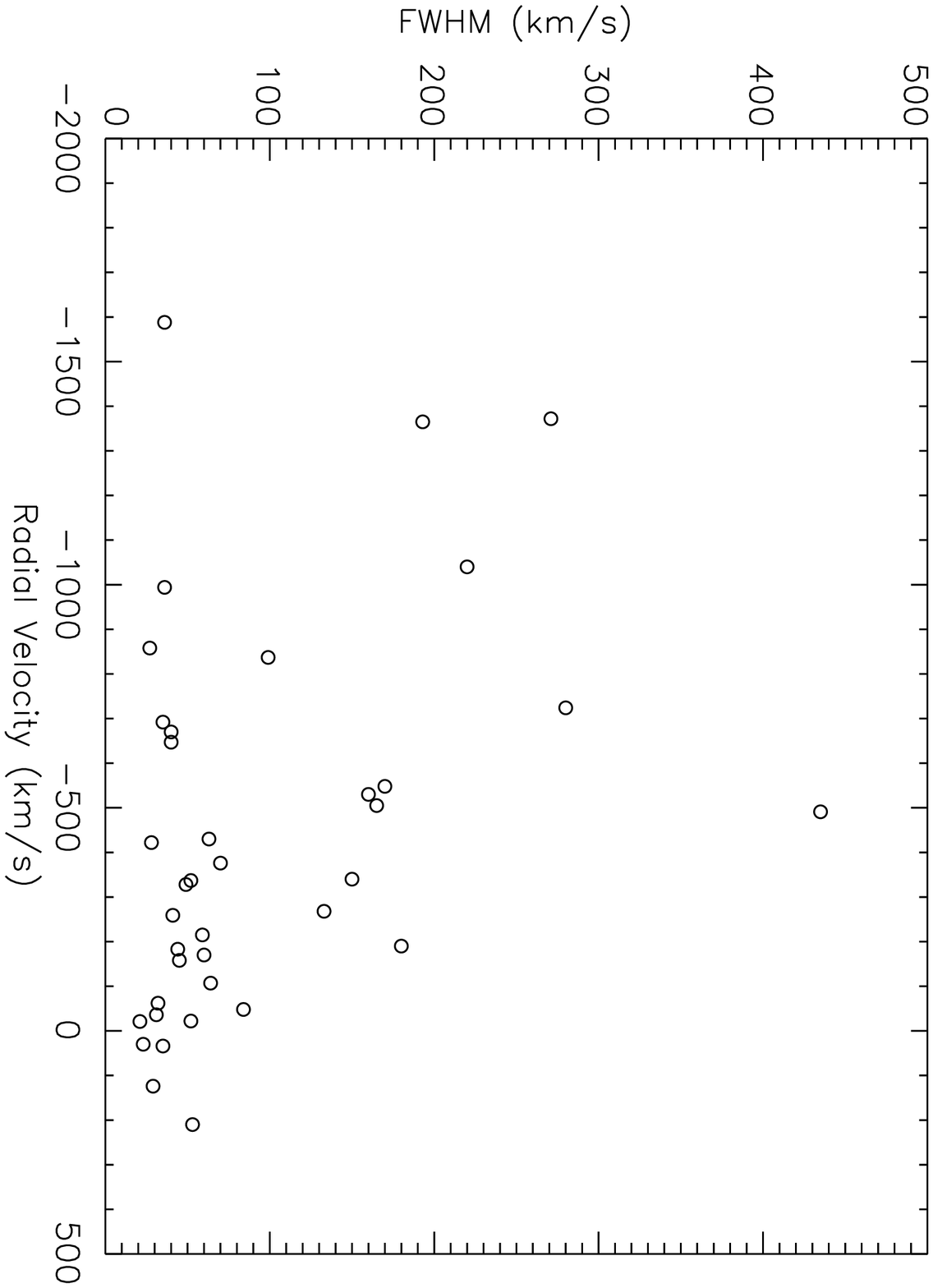]{FWHM vs. radial-velocity centroid for the UV absorbers.}

\figcaption[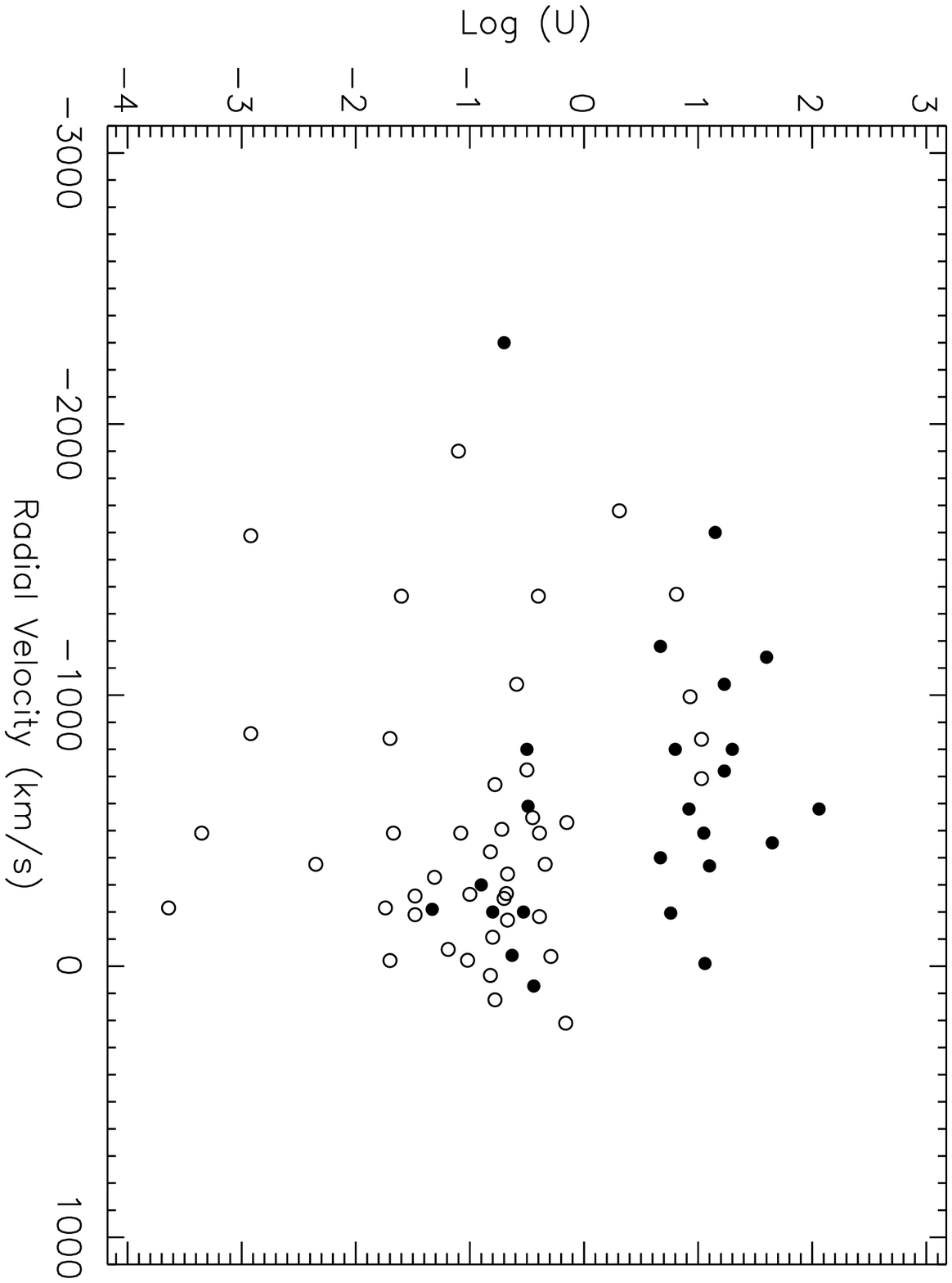]{Ionization parameter vs. radial-velocity centroid for the UV
(open circles) and X-ray (filled circles) absorbers.}

\figcaption[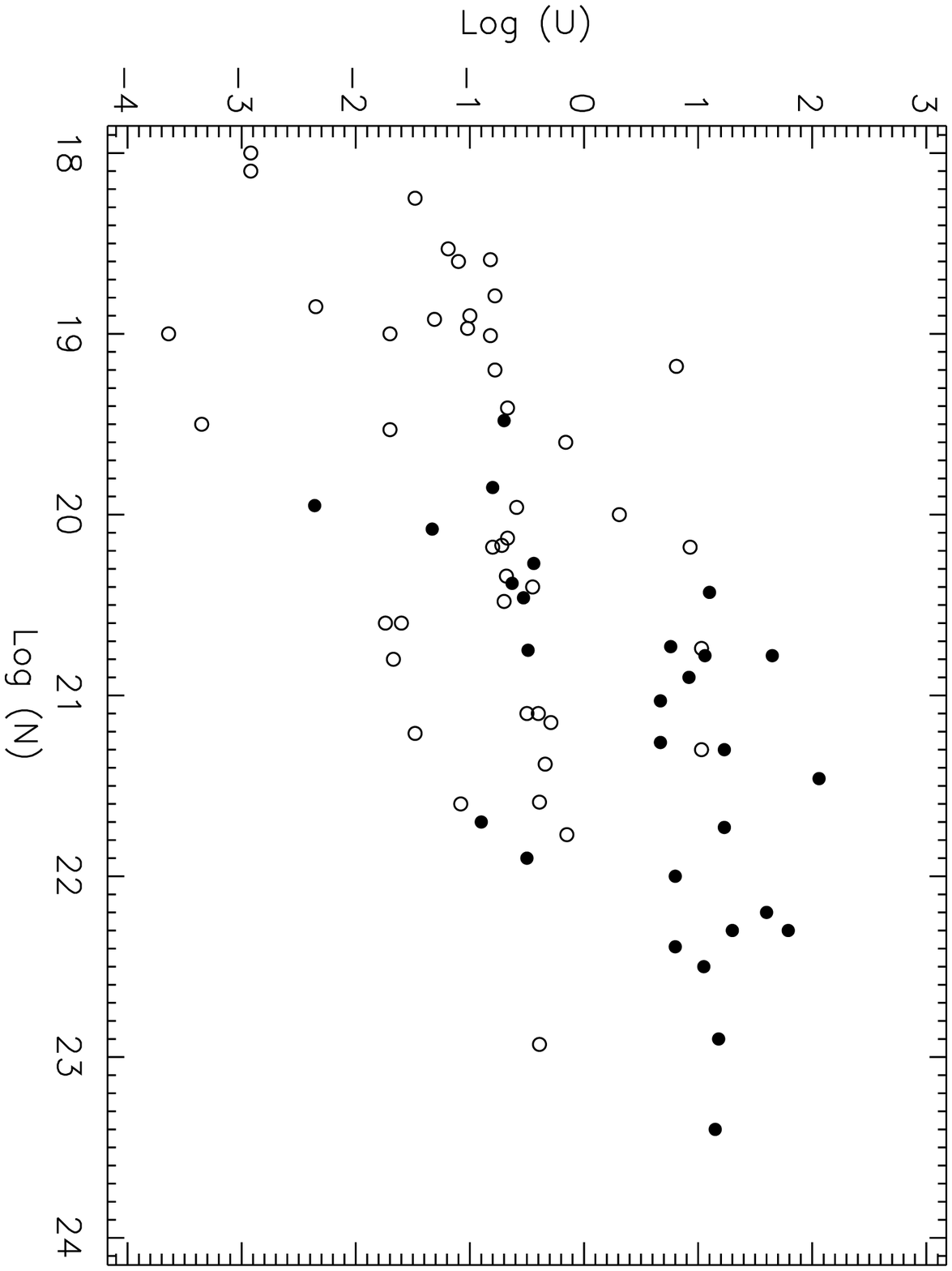]{Ionization parameter vs. column density (cm$^{-2}$) for the
UV (open circles) and X-ray (filled circles) absorbers.}

\figcaption[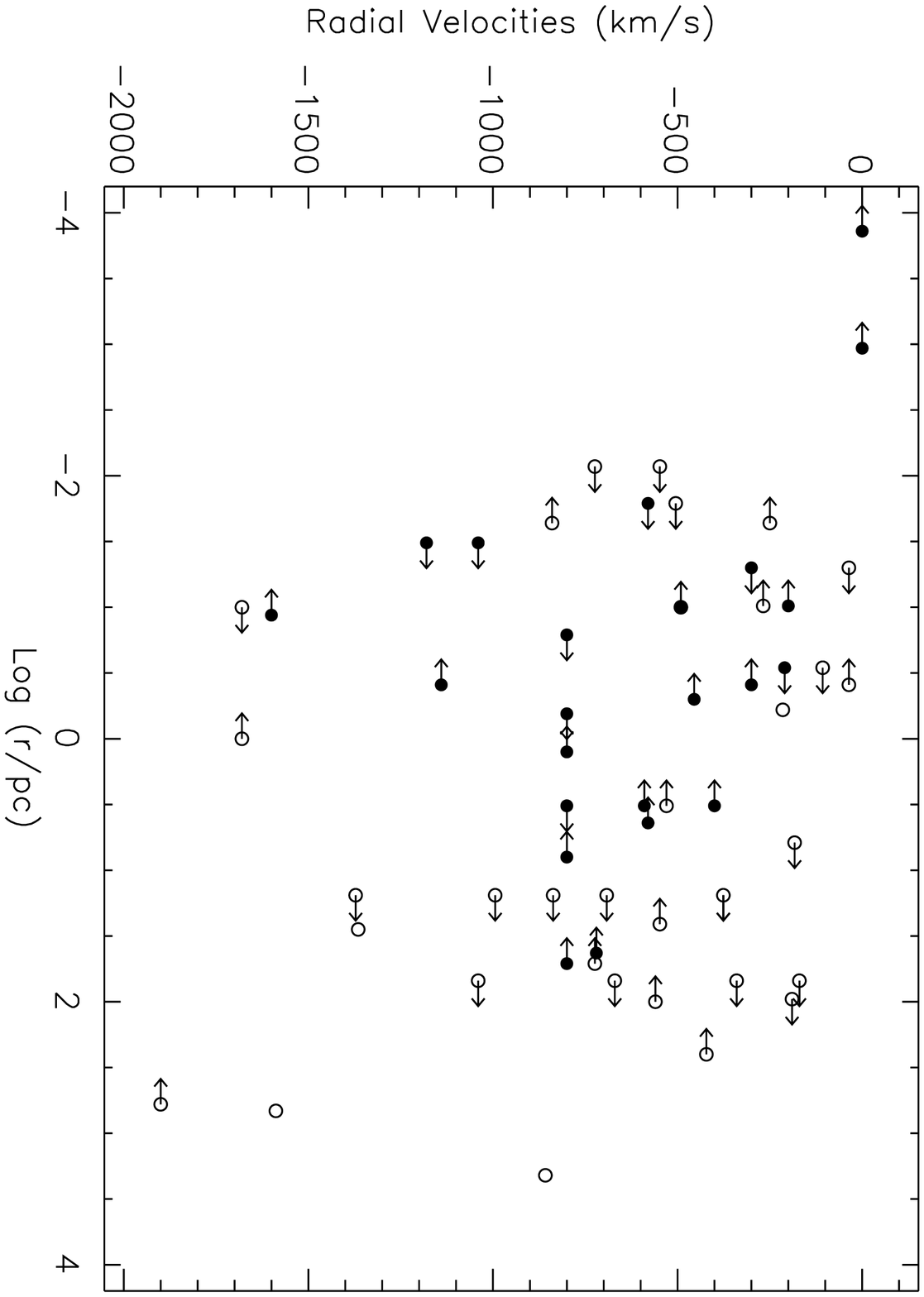]{Radial location (or limit) vs. radial velocity
centroid (km s$^{-1}$) for the UV (open circles) and X-ray (filled circles)
absorbers.}

\figcaption[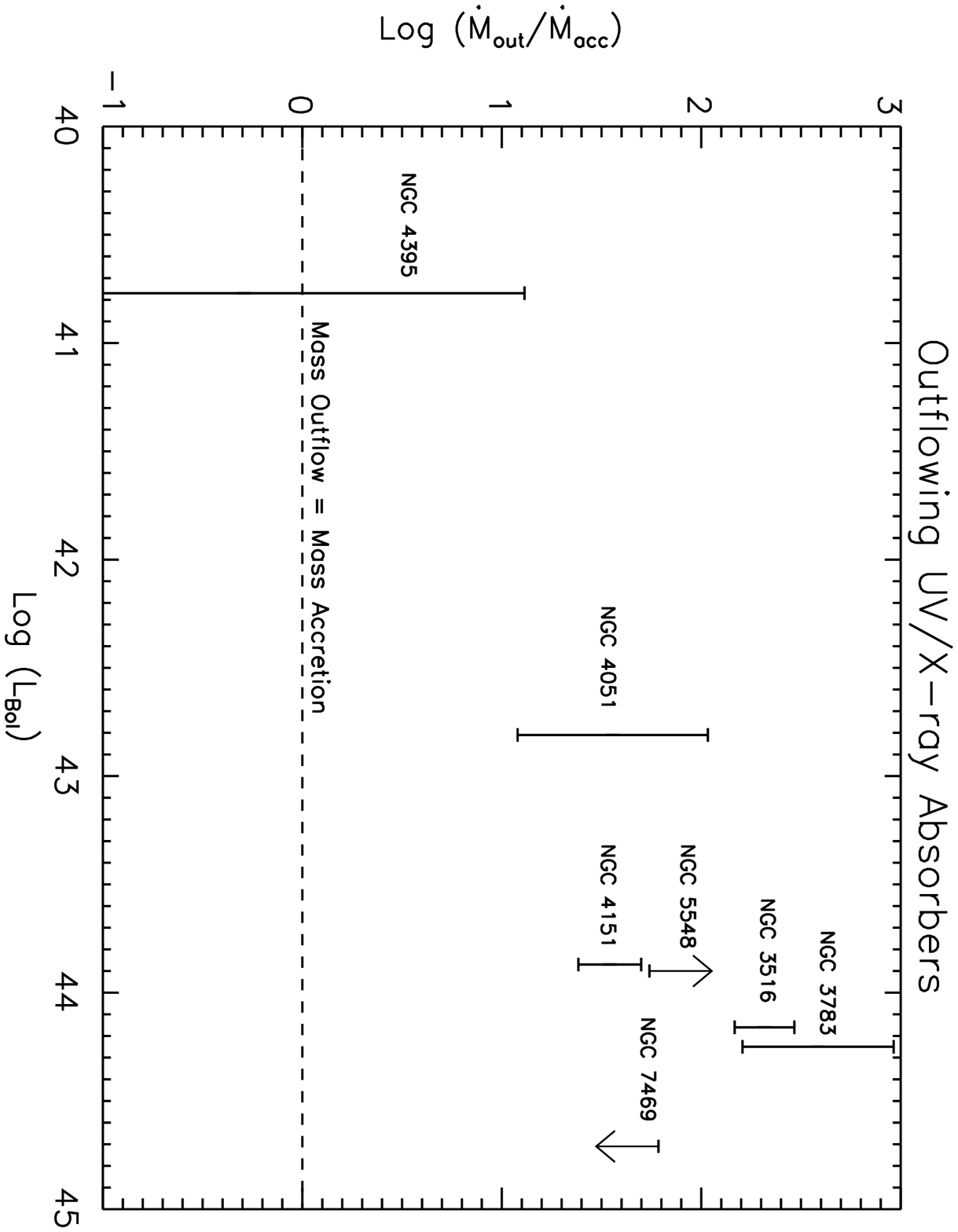]{Ratio of total mass outflow rate to inferred accretion rate
as a function of bolometric luminosity, based on the minimum and maximum
values given in Table 2.}

\figcaption[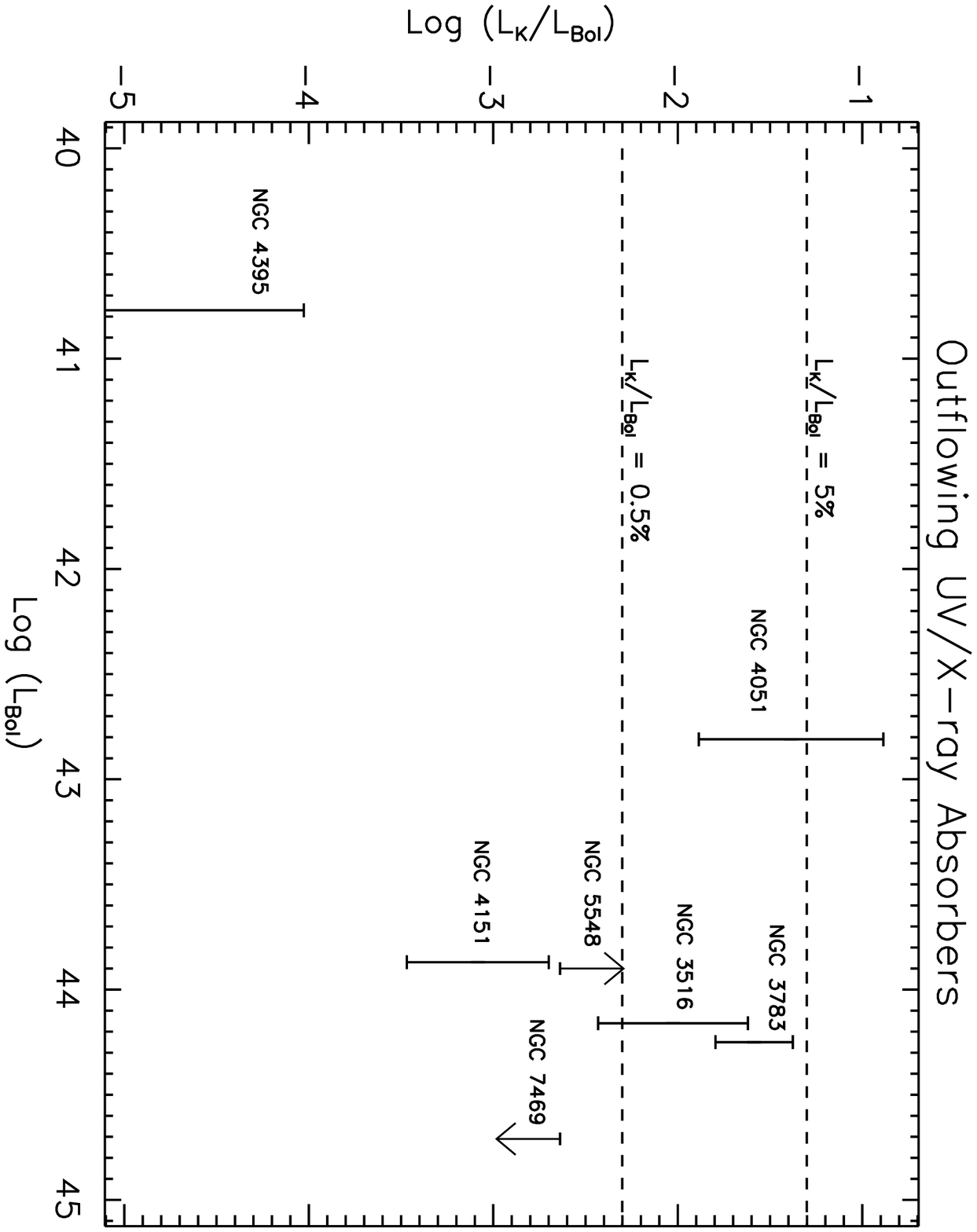]{Ratio of total kinetic luminosity to bolometric luminosity
as a function of bolometric luminosity, based on the minimum and maximum
values given in Table 2.}

\clearpage
\begin{figure}
\plotone{f1.eps}
\\Fig.~1.
\end{figure}

\clearpage
\begin{figure}
\plotone{f2.eps}
\\Fig.~2.
\end{figure}

\clearpage
\begin{figure}
\plotone{f3.eps}
\\Fig.~3.
\end{figure}

\clearpage
\begin{figure}
\plotone{f4.eps}
\\Fig.~4.
\end{figure}

\clearpage
\begin{figure}
\plotone{f5.eps}
\\Fig.~5.
\end{figure}

\clearpage
\begin{figure}
\plotone{f6.eps}
\\Fig.~6.
\end{figure}

\end{document}